# All-epitaxial self-assembly of silicon color centers confined within atomically thin two-dimensional layers using ultra-low temperature epitaxy

Johannes Aberl[1,*], Enrique Prado Navarrete[1], Merve Karaman[1], Diego Haya Enriquez[1], Christoph Wilflingseder[1], Andreas Salomon[1], Daniel Primetzhofer[2], Markus Andreas Schubert[3], Giovanni Capellini[3,4], Thomas Fromherz[1], Peter Deák[5], Péter Udvarhelyi[5,6], Li Song[5], Ádám Gali[5,6,7], Moritz Brehm[1,*]

[1] *Institute of Semiconductor and Solid State Physics, Johannes Kepler University, Altenberger Straße 69, Linz 4040, Austria*

[2] *Department of Physics and Astronomy, Uppsala University, Box 516, 75120 Uppsala, Sweden*

[3] *IHP—Leibniz-Institut für innovative Mikroelektronik, Frankfurt (Oder), Germany*

[4] *Dipartimento di Scienze, Universita degli Studi Roma Tre, Rome, Italy*

[5] *HUN-REN Wigner Research Centre for Physics, P.O. Box 49, H-1525, Budapest, Hungary*

[6] *Department of Atomic Physics, Institute of Physics, Budapest University of Technology and Economics, Műegyetem rakpart 3., Budapest H-1111, Hungary*

[7] *MTA-WFK Lendület "Momentum" Semiconductor Nanostructures Research Group, P.O. Box 49, H-1525, Budapest, Hungary*

\* Johannes.aberl@jku.at
\* moritz.brehm@jku.at

**Abstract**

**Silicon-based color-centers (SiCCs) have recently emerged as quantum-light sources that can be combined with telecom-range Si Photonics platforms. Unfortunately, using current SiCC fabrication, deterministic control over the vertical emitter position is impossible due to ion-implantation's stochastic nature. To overcome this bottleneck towards high-yield integration, we demonstrate a radically innovative creation method for various SiCCs, solely relying on epitaxial growth of Si and C-doped Si at atypically-low temperatures in a ultra-clean growth environment. These telecom emitters can be confined within sub-1nm thick layers embedded at arbitrary vertical positions within a highly crystalline Si matrix. Tuning growth conditions and doping, different SiCC types, e.g., W-centers, T-centers, G-centers, or derivatives like G'-centers can be created, which are particularly promising as Si-based single-photon sources and spin-photon interfaces. The zero-phonon emission from G'-centers can be conveniently tuned by the C-concentration, leading to a systematic wavelength shift and linewidth narrowing towards low emitter densities.**

**Introduction**

Since the early days of silicon (Si) electronics, the multitude of different Si color centers (SiCCs) induced by ion-implantation and other treatments in semiconductor technology have been studied extensively.[1] The investigations mainly focused on their structural properties, aiming to use SiCCs as fingerprints to probe material quality. However, recent works revealed that isolated, single SiCCs could serve as on-demand sources of single telecom photons[2-6] and light-matter interfaces[7-9] –fundamental building blocks for advanced quantum technologies including quantum communication, and computation.[10] This huge potential has motivated an ever-growing scientific community to invesitigate a variety of newly-discovered optically-active single SiCCs.[4] The most prominent are related to carbon-Si point defects, carbon-hydrogen-Si, oxygen-Si complexes, or Si self-interstitials.[1-9,11-22] For many SiCCs, the research regarding their fundamental structural, electronic[11], and optical properties, and benefits and drawbacks of host modifications[21,23], resilience against nearby lattice imperfections is still in its infancy. Despite the lack of yet optimized manufacturing schemes for high-quality SiCCs, some technological milestones, such as electrically-pumped emission from SiCC ensembles[12-14] or their integration into photonic waveguides[18,19] and resonators[15-17] have already been demonstrated. Considering photonic integration, these quantum emitters can significantly benefit from the Si-based photonic integrated circuit technology[15-19], where they could be monolithically integrated, unlike other proposed solid-state quantum light sources.

Unfortunately, all these defects result from stochastic crystal lattice damage due to ion-implantation at typical energies ranging from ~20keV-150keV.[2-9,11-13] This dependency on ion-implantation poses the major bottleneck towards deterministic high-yield photonic integration, as implantation leads to extended ion distribution profiles and, thus, insufficient control over the SiCC formation process with typical depth variations of more than hundred nanometers. Figure 1a shows the mean projected range and two times the vertical straggling ($\sigma$) obtained from the transport of ions in matter (TRIM) simulations[24] for three relevant ion species ($Si^+$, $C^+$, and $H^+$, respectively) implanted into Si. Within a vertical range of at least 100 to 200nm, the emitter formation occurs stochastically and cannot be controlled, even if only single ions are implanted. Figure 1b depicts a common fabrication scheme for SiCCs on silicon-on-insulator (SOI) substrates. The implantation of $C^+$ typically leads to the formation of so-called G-centers, point defect complexes formed by an interstitial Si atom bound to two substitutional carbon atoms.[1] Typically, high-energy $C^+$-implantation is followed by thermal annealing and an optional $H^+$-implantation. An implantation energy of ~35keV for $C^+$ corresponds to a projected range of 110 nm, i.e., the maximum of the defect formation occurs at the center of a 220nm thick SOI layer and is accompanied by a broad vertical emitter distribution (Fig. 1b).

This random formation process limits deterministic photonic and -device integration at large scales, which asks for an optimized and reproducible overlap of the vertical emitter position with the field maxima of the photonic modes in waveguides and resonators. Implantation at atypically-low energies (≤ 5keV) could minimize the implantation damage to a depth window of just a few nanometers.[25] However, the resulting vicinity of the emitter to the surface is non-advantageous regarding parasitic surface states, poor emitter/mode overlap, and spectral diffusion-related issues for indistinguishability. Post-implantation overgrowth will not be able to eliminate this interface, and the low thermal budget of SiCC during epilayer preparation will always be a major concern. This is particularly true considering the sensitivity of quantum emitters on their local matrix environment that can limit the SiCC efficiency (Aberl et al, in preparation).

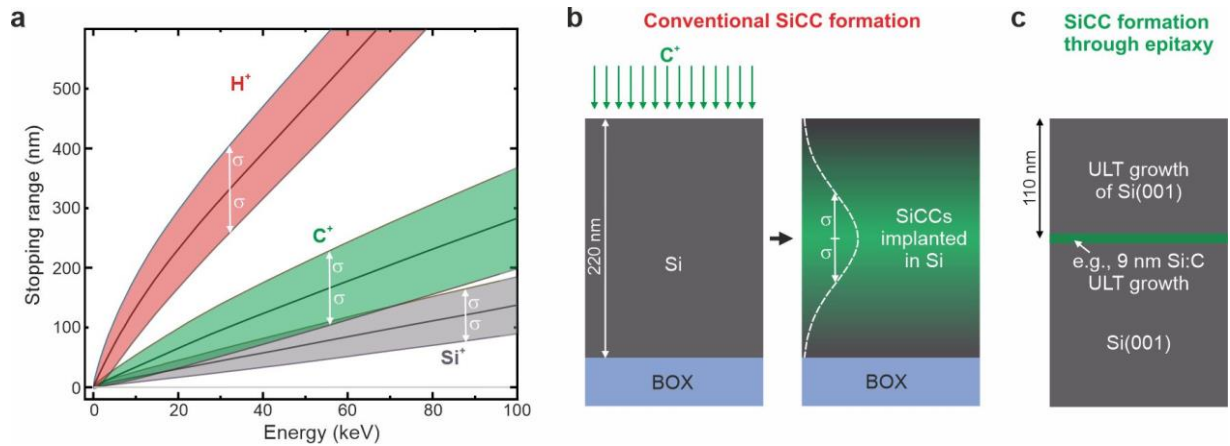

*Fig. 1| Intrinsic bottleneck of using ion-implantation for color center formation. a Simulated stopping range of $Si^+$, $C^+$ and protons in crystalline Si(001) for different common implantation energies. b Common fabrication scheme for Si-color centers (SiCC) consisting of ion-implantation (here $C^+$) and annealing, leading to a stochastic distribution of emitters in the SOI device layer. c SiCC formation through molecular beam epitaxy growth at ultra-low sample temperatures (200°C-300°C). Efficient confinement of the emitter position down to the nanoscale through the growth of thin Si:C layers, overgrown with Si at ULT conditions.*

Here, we introduce an entirely different way by creating SiCCs without ion-implantation but, instead, using SiCC self-assembly during epitaxial growth at unconventionally low temperatures (ULT-growth). This novel approach provides decisive benefits, similar to quantum dot emitters, in confining the vertical emitter position within the epilayer to the nanoscale (Fig. 1c). As proof of concept, we demonstrate the confinement of the SiCCs within a layer of one Si lattice constant (5.4Å) thickness that can be perfectly aligned in the vertical center of a 220nm thick SOI layer.

## Results

**Vertical confinement of Si color centers**

We first demonstrate the selective epitaxial creation of SiCCs within sub-10nm thick layers and at deterministically chosen depths under the sample surface. For all applications, the thin SiCC layers must be overgrown to ensure, e.g., optimal emitter/cavity-mode overlap and provide the necessary vertical separation between emitters and parasitic surface states. A typical growth protocol is depicted in Fig. S1, Supplementary Material. A high-quality Si buffer layer grown at high growth temperature ($T_G$>500°C) separates the SiCC-layer from the initial substrate surface. Then, we grew 9nm of Si or Si:C at a $T_G$=200°C. This $T_G$ was kept constant for Si:C deposition throughout this work. The low $T_G$ limits the growth kinetics and allows for the epitaxial formation of SiCCs that are subsequently overgrown with Si at capping-layer growth temperatures $T_{cap}$s varying from 200°C to 310°C. We changed $T_{cap}$, to find an ideal trade-off between the Si matrix quality and the thermal budget, which must not negatively affect the SiCC properties, as they collectively annihilate above ~300°C.[1] In Fig. 2, the Si capping layers thickness was ~105 nm to mimic an emitter position precisely in the middle of a SOI-220nm substrate. To unambiguously trace the PL's origin to the SiCC layer, we studied the PL response of pure Si reference layers grown at ULT, see Fig. 2.

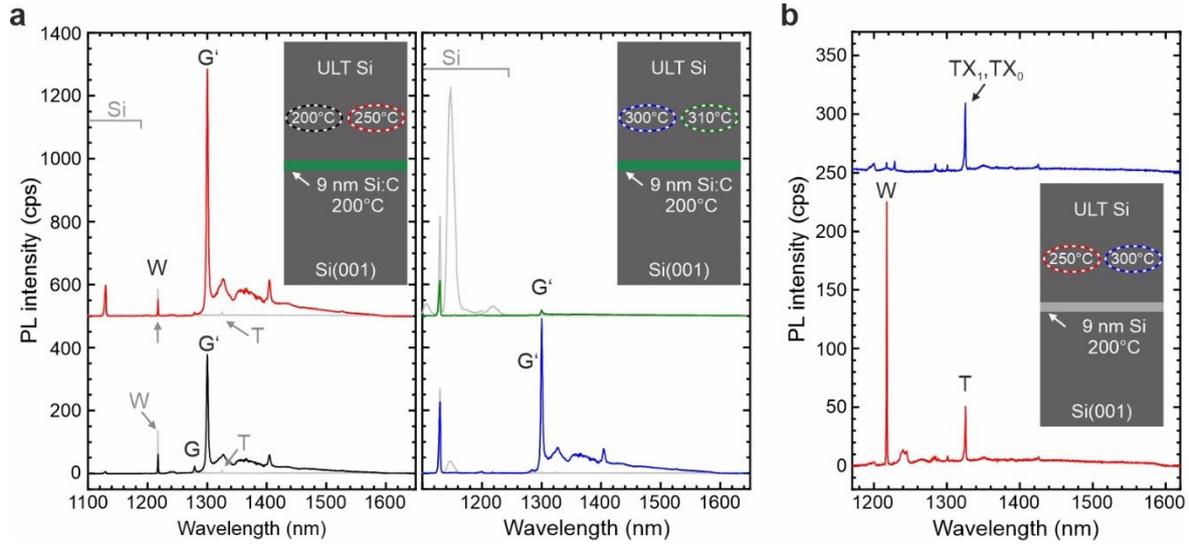

***Fig. 2| Photoluminescence spectra of Si color centers formed within nanoconfined layers.***
***a*** *Deposition of 9 nm Si:C at* $T_G$ *= 200°C and overgrown with Si at* $T_{cap}$ *= 200°C (black spectrum),* $T_{cap}$ *= 250°C (red spectrum),* $T_{cap}$ *= 300°C (blue spectrum), and* $T_{cap}$ *= 310°C (green spectrum). Insets show the respective fabrication scheme. Grey PL spectra originate from pure Si reference samples, grown at the respective* $T_{cap}$*s of 200°C, 250C°C, 300°C and 310°C. Predominant formation of G' color centers in the 9 nm Si:C layer and W-centers in the Si capping layer at* $T_{cap}$ *≤ 250°C.* ***b*** *Predominant self-assembly of W-centers (red spectrum) and T-centers (blue spectrum) through the growth of 9 nm Si at* $T_G$ *= 200°C followed by Si overgrowth at* $T_{cap}$ *= 250°C and* $T_{cap}$ *= 300°C, respectively. Some spectra are vertically shifted for clarity.*

Figure 2a depicts the PL emission from 9nm thick Si:C layers (C=3.8·10$^{19}$cm$^{-3}$) grown at $T_G$=200°C and capped with pure Si at $T_{cap}$=200°C, 250°C, 300°C, and 310°C. The superimposed gray spectra in each plot correspond to reference spectra for which the low-temperature Si:C layer was replaced by a Si layer. For the Si references grown at $T_{cap}$≤250°C, spectrally narrow emission at 1220nm and 1325nm is observed (grey arrows), corresponding to W-center and T-center zero phonon lines (ZPL).[1] For $T_{cap}$≥300°C, the PL spectra of the reference samples in Fig. 2a indicate an excellent crystal quality, as evidenced by two findings. First, for $T_{cap}$=300°C, only faint PL emission lines below the Si bandgap are visible that vanish at $T_{cap}$=310°C. Second, at $T_{cap}$≥300°C, we also observe a broad electron-hole droplet (EHD) related emission from the Si matrix layer at a wavelength of 1150 nm, a sign of high crystal quality and low density of non-radiative recombination channels. Note that for $T_{cap}$≤250°C, for which W and T-center PL is observed in the reference samples, no EHD emission from Si exists. We also note that for the samples containing the Si:C layer, no EHD emission is observed, as carriers efficiently recombine within the SiCC layer. The excellent crystalline quality of the Si capping layer at $T_{cap}$=300°C is confirmed by high-resolution cross-sectional TEM, see Fig. S2, of the Supplementary Material.

In Fig. 2a, the shapes of the grey reference PL signals starkly contrast the signal from the samples containing the C-doped layer. For all C-doped samples, we observe a prominent ZPL line at 1299.8nm (953.8meV). We note that the observed ZPL wavelength deviates from that of G-center at ~1278 nm (Ref. 1), but its observed Debye-Waller factor at 17.9% is very similar to that of a G-center ensemble.[4] Like the G-center, the here-found emission center consists of a ZPL, a broad phonon side band at higher wavelengths, and a local phonon mode that is shifted by about 71 meV to lower energies (Fig. S3a). Due to the similarities in their spectral shape, we label the emitter as G'. Indeed, the here-found G'-center might be one of the G-center-related lines found in an earlier work by Davies *et al.* (Ref. 26).

Our *ab initio* calculations imply that the G'-center consists three carbon atoms where a substitutional carbon sits next to the atomic configuration of the G-center which well reproduces the characteristic features of the PL spectra: red-shift both in the ZPL and the prominent localized vibration peak near the ZPL (also-called 'E' peak[1]) in the G'-center when compared to the G-center (see Supplementary Materials for details). We note that our carbon-source contains molecules with three carbon atoms that might be built in as a unit to the Si crystal, which may facilitate the dominant formation of G'-center.

For the SiCC layer with $T_{cap}$≤250°C, the W-center emission is weaker than for the corresponding reference sample ($T_{cap}$≤250°C), pointing to a dominant carrier capture within the thin Si:C layer. At $T_{cap}$=300°C, blue spectrum in Fig. 2a, we still observe a strong G'-center signal, while the background W-center signal from the Si matrix vanished, strongly indicating that the SiCC emission is unambiguously dominantly originating from the thin ULT Si:C region. Segregation of C at $T_G$ of Si:C nanolayer and Si capping layer can be excluded.[27] The ratio of the integrated G'-ZPL intensity of the sample with C=3.8×10$^{19}$cm$^{-3}$ and $T_{cap}$=300°C relative to the reference sample without C-deposition is as high as 310:1. However, we note that the thermal budget of the SiCCs is critical since the thermal annealing induced by overgrowth at a $T_{cap}$=310°C induces a strong quenching of the G'-center emission (green spectrum in Fig. 2a).

Results of excited state lifetime measurements obtained by time-correlated single photon counting for ensembles of G'-centers, overgrown at $T_{cap}$=200°C and 300°C, are shown in Fig. S3b of the Supplementary Material. The PL lifetimes are ~7 ns for both samples, a value close to the one observed for G-centers[2,6,18], and significantly shorter as compared to the lifetimes observed for W- and T-centers.[5,7]

Nevertheless, the occurrence of W- and T-centers in the Si reference samples grown at $T_{cap}$ ≤ 250°C offers the further possibility to select the emitter type within a nanolayer. By growing a 9 nm thick Si layer at $T_G$ = 200°C and overgrowing it with Si at $T_{cap}$ = 250°C and $T_{cap}$ = 300°C, respectively, W-centers and T-centers[1] can be predominantly grown in the nanolayer, Fig. 2b. The W-centers created recent interest as single-photon emitting SiCCs,[5] while the T-center, a Si-C-H point defect complex, is particularly promising due to its long spin lifetimes and spin-selective optical transitions.[21]

**Influence of C-doping and layer thickness on SiCC emission**

Previously, we focused on a high C-concentration (3.8×10$^{19}$cm$^{-3}$) and a Si:C layer thickness of 9nm to determine the SiCC's vertical selectivity. Next, we focus on the influence of the C-concentration and the Si:C layer thickness on the PL emission. Figure 3a depicts the impact of a decreasing C concentration (C=5×10$^{20}$cm$^{-3}$ to 2.2×10$^{17}$cm$^{-3}$) in a 9nm thick Si:C film that was subsequently capped with ~105nm of Si, grown at $T_{cap}$=300°C. For low C-concentrations of 2.2×10$^{17}$cm$^{-3}$, the spectrum is dominated by the ZPL of the G'-center at 1300.8 nm (953.2 meV) (Figs. 3a,e,f) with a detectable FWHM of less than 0.4nm (< 300µeV). With increasing C-concentration, the PL intensity of the G'-ZPL emission increases (Figs. 3a,c). At the same time, the ZPL emission blue-shifts (Figs. 3a,e), and the FWHM increases, see Fig. 3f. For very high C-concentration of 5×10$^{20}$cm$^{-3}$, the spectrum is dominated by the emission of the phonon-related sideband, exhibiting only a broad zero phonon peak at about 1295nm (Fig. 3a,f). At the lowest investigated C-concentrations, the spectral signature of T-centers is observed at wavelengths around 1325nm (TX$_0$ at 1326nm and TX$_1$ at 1323.5nm)[8].

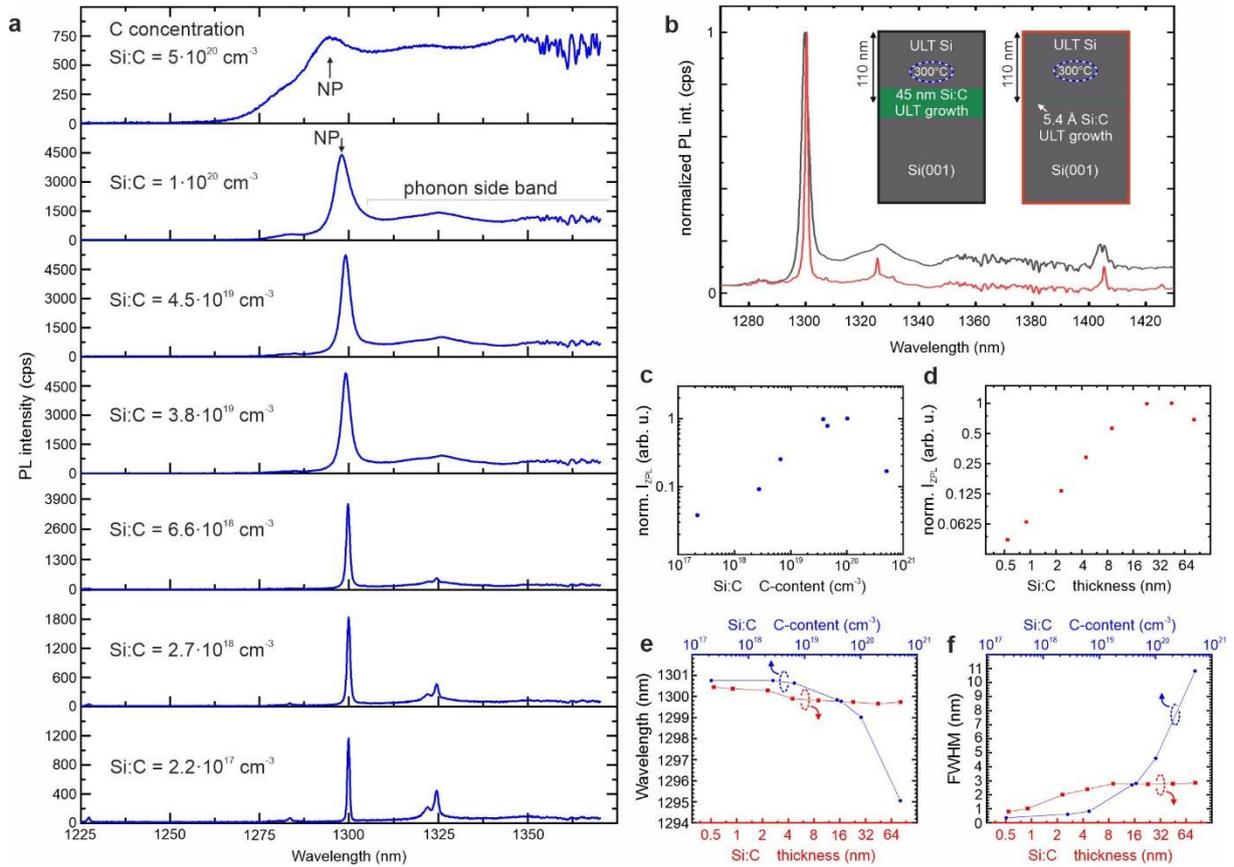

*Fig. 3| a Photoluminescence spectra of Si color centers. SiCCS, formed upon deposition of a 9 nm thick C-doped Si layer with various C concentrations, deposited at $T_G$ = 200°C and overgrown at $T_{cap}$ = 300°C. b Si:C layers of 45 nm thickness (black spectrum) and 0.5 nm thickness (red spectrum), deposited at $T_G$ = 200°C and capped at $T_{cap}$ = 300°C. A C-concentration of $3.8·10^{19}$ $cm^{-3}$ was used. Insets show a scheme of the respective sample structure. c Influence of the C-concentration in the Si:C layer on the normalized integrated PL intensity of the zero-phonon line ($I_{ZPL}$). d Influence of the Si:C layer thickness on $I_{ZPL}$. e Influence of the Si:C layer thickness and C-concentration in the Si:C layer on the emission wavelength of the zero-phonon line of the G' center. f Influence of the Si:C layer thickness and C-concentration in the Si:C layer on the full-width-at-half-maximum (FWHM) of the zero-phonon line of the G' center.*

Figure 3b depicts the differences in PL emission from SiCC layers with thickness of 45nm and 5.4Å. Both spectra are normalized to the maximum of their G' ZPL emission. The comparison indicates a narrowing of the ZPL line and an increasing Debye-Waller factor with decreasing Si:C layer thickness. The inset in Fig. 3b shows the respective sample layout, with both Si:C layers being doped with C to a concentration of $3.8×10^{19}cm^{-3}$. The capping layer thickness ($T_{cap}$=300°C) was adjusted so that the middle of the Si:C layer was exactly 110nm below the sample surface. The dependence of the ZPL intensity, transition wavelength, and linewidth on the Si:C layer thickness is shown more systematically in Figs. 3d-f. Continuously increasing the Si:C layer thickness from 0.5nm to ~30nm, the PL intensity of the G'-ZPL increases linearly with increasing film thickness (Fig. 3d), pointing to an excellent scalability of the SiCC densities. For larger thickness, (≥45nm), the emission intensity saturates and even slightly decreases. We assign this decrease to the thinner Si capping layer for thicker Si:C layers and, thus to a decreased in-situ annealing time, associated with the Si capping layer growth at $T_{cap}$ = 300°C. Comparing the emission intensities in Fig. 2a, it is evident that moderate annealing at 250°C-300°C increases the emission intensity from SiCCs. At the same time, the emission wavelength of the G'-ZPL slightly increases with decreasing thickness (Fig. 3e), while the FWHM of the G'-ZPL decreases.

The characteristics of these interstitial-related SiCCs significantly constrain the process parameter window for in-growth defect engineering. The relatively-low thermal budget that can be tolerated by SiCCs before annihilation impose the use of ULT epitaxy, with growth temperatures ($T_G$ and $T_{Cap}$) limited to <300°C. Earlier works in proton implanted samples found that for traditional G-centers, disintegration starts already at ~175°C for a 30-minute annealing step in argon atmosphere, while for W-centers, this temperature is ~270°C.[28] Therefore, a suitable $T_G$-balance between SiCC generation and preservation or emitter reduction during Si capping and annealing has to be considered. Here, we find that growing SiCCs at $T_G$=200°C and embedding them in Si grown at $T_{cap}$=300°C minimizes the damage in the Si matrix. A finer adjustment of $T_G$s will likely lead to further optimization of the emitter/matrix quality. We also emphasize the need for ultra-low chamber pressures *during growth*. At growth temperatures <350°C, residual gases are not efficiently desorbed from the Si surface,[29] i.e., foreign residual atoms that constitute the growth chamber pressure (background plus pressure increase due to hot sources) are also incorporated. The relation between growth pressure and rate of residual gas impingement[30] is shown in Fig. S1 in the Supplementary Material. These foreign atoms can form detrimental defects that, in the worst case, disturb SiCC emission through the generation of (non)-radiative recombination channels and provide an unfavorably noisy matrix environment for quantum emitters. Here, we find a spurious PL response from T-centers (C-H-related defects), despite excellent growth pressures in the low $10^{-10}$ mbar range (see Fig. S1, Supplementary Material). However, this finding guides future experiments for the creation of high-quality isolated T-centers.

We note that many different point-defect centers are known to exist in Si.[1-9,11-22] However, due to the novelty of their application potential in quantum communication, it can be expected that only a fraction of defect-centers with optimized quantum properties are known to date. Their properties will depend on factors like, e.g., types of foreign atoms at the point defect site, strain, matrix composition, etc. Here, we demonstrate the epitaxial creation of various vertically confined emitters, such as G, G'-, W-, and T-centers that are, when isolated, promising candidates for quantum photonic applications. In particular, our *ab initio* calculations show (see Supplementary Materials) that G'-center should exhibit a similar optically detected magnetic resonance (ODMR) signal as its G-center counterpart.[31] Thus, the G'-centers have a great potential to act as a quantum memory together with emission in the telecom O-band.

## Discussion

This first demonstration of purely epitaxial growth of self-assembled SiCCs marks the critical initial step towards deterministic vertical position control in the nanometer range. As a proof of concept, we have confined the G' emitters within an only one unit cell thick layer (5.4Å). Additionally, omitting the overall damaging impact of ion-implantation promises emitter generation within a matrix material that can be electronically more "quiet", important for any kind of quantum application, especially considering the detrimental influence of spectral diffusion on the indistinguishability of quantum emitters.[32] In this work, these ground-breaking methods have already been shown to be excellent for forming different SiCC types, and a successful extension to even more types can be expected. We point out that this initial work was targeted towards a proof-of-principle and a vast parameter space for optimizing the SiCCs with respect to their quantum optical benchmarks remains untapped.

As the next milestone to achieve laterally separated single SiCCs, a systematic reduction of C-concentrations, layer thicknesses, and annealing or a combination of these has to be addressed.

Sample annealing to dilute the emitter density, performed in-situ in the MBE's UHV environment after Si matrix growth is a logical next step. Furthermore, conventional ex-situ annealing methods as employed for ion-implanted SiCCs can be used. Additionally, templated local epitaxy schemes for fabricating self-assembled SiCCs can be envisioned to control the lateral emitter position. Until then, deterministic processing techniques that have been developed for randomly nucleated group-III-V QDs, like in-situ optical lithography[33], cathodoluminescence lithography[34], and two-color PL imaging[35,36] can be applied to self-assembled SiCCs.

The use of *in-situ* defect incorporation and growth of the host Si crystal has several further advantages, providing various opportunities. It is viable to grow Si on top of SOI. Therefore all resonator structures (microdisks, photonic crystals, bulleyes, waveguides, etc.) can be conveniently fabricated top-down. In contrast to other approaches,[37] our present method can straightforwardly position SiCCs deterministically at vertical positions, where the photonic mode has a maximum. Based on previous results, we envision that Q-factors of $10^5$-$10^6$ can be achieved in, e.g., photonic crystal cavities[38,39]. This is definitely of great importance for T-centers for which the brightness and the quantum yield is not sufficiently high for pursuing quantum optical protocols on single defects.[7] In addition, we envision deterministic strain control for SiCCs with our scheme. Strain control on SiCCs is of prime importance as the spin levels[7,40] and the position and polarization of the ZPL emission are greatly sensitive to the strain field in Si.[41] By adding Ge to the Si crystal, $Si_{1-x}Ge_x$ heterostructures can be readily grown where the variation of composition *x* can be used to tune the strength of the desired strain field. Indeed, our ULT-MBE growth applied to SiGe heteroepitaxy[42] also allows for the creation of high-quality planar top-down platforms for advanced nanoelectronics devices.[43,44] Furthermore, the proposed growth technique can be employed to realize long coherence times for the electronic spins of SiCC qubits and engineer the ancilla qubits around the SiCC qubits such as T-center. We note that once the ODMR signal from a single G-center or G'-center is observed, it could step in as a strong contender in the field of quantum communication, where nuclear spin quantum memories are essential in realizing quantum repeaters. By using $^{28}$Si enriched growth of SiCC layer and its immediate surroundings, the coherence times of SiCC may reach milliseconds[8,45] where the nuclear registers can be engineered at the target distance by adding natural Si or any other artificial isotope contribution by design. This can be used to tune the coupling of the electron spin and nuclear spin so as to increase the rate of quantum information processing between the nuclear spin register and the electron spin. Furthermore, the quasi-two-dimensional layer of $I$=1/2 nuclear spins can be used in quantum simulation protocols[46] or to build up a quantum chip when the individual nuclear spins can be addressed and manipulated with the combination of microwave and radiofrequency pulses.[47] Finally, vertical p-i-n junctions can be readily fabricated by this growth technique where the SiCCs are engineered in the intrinsic region to avoid charge fluctuations under illumination of SiCCs[48] and apply controllable Stark-shift on the SiCC's ZPL or to realize photoelectric readout of spins.[49-52]

In summary, we expect that the here presented ULT-growth approach for creating vertically controlled Si color centers can be a starting point for exploring the untapped potential of telecom SiCC quantum emitters and qubits with highly homogeneous environment because of the non-invasive creation of SiCCs and the full control of optical, electrical, strain and spin environments around SiCCs. Furthermore, this fully integrated quantum electro-optics device can be readily connected to optical fibers by tapering[53] to the silicon nanobeams or nanopillars in order to maximize the photon output and integrate this to commercial optics devices.

## Methods

**Epitaxial growth**

All samples were grown on 17.5 mm x 17.5 mm float zone (FZ) Si(001) substrates ($R_0>5000$ Ωcm) cut from a 4-inch wafer. Subsequently, the protective photoresist applied for cutting was removed within a precleaning procedure based on solvents (acetone and methanol) and a UV-ozone surface treatment. Before the growth, the substrates were prepared using a Radio Cooperation of America (RCA) cleaning and were dipped in diluted hydrofluoric acid (HF1%) to remove the native oxide before being introduced into the load lock chamber. All samples were degassed at 700°C for 15 minutes, followed by a conditioning step at 450°C for 30 minutes prior to the growth. A 75.5 nm thick Si buffer layer was grown at a $T_G$, ramped from 650°C to 600°C. For all samples containing a layer of C-doped Si, $T_G$ was ramped down during a growth interrupt to 200°C, the growth temperatures of all SiCCs (see Fig. S1, supplementary material). To investigate the $T_{cap}$ dependence, we deposited 9 nm thick Si:C layers with a nominal Si growth rate of 0.5 Å/s and C-concentrations of $3.8 \cdot 10^{19}$ cm$^{-3}$ at $T_G$ = 200°C. Hereafter, the substrate temperature was ramped to the respective Si $T_{cap}$ of 310°C, 300°C, 250°C or 200°C, and a 104.5 nm thick Si capping layer was grown at a growth rate of 0.75 Å/s. The respective Si capping layers were deposited directly onto the Si buffer layers for the reference samples. For the investigation of the C-concentration dependence, the Si:C layers' thickness was kept constant at 9 nm, while the C concentration varied from $2.2 \cdot 10^{17}$ cm$^{-3}$ to $5.0 \cdot 10^{20}$ cm$^{-3}$. The carbon deposition rates were calibrated using secondary-ion mass spectrometry (SIMS) experiments of calibration layers. For the investigation of the Si:C layer thickness dependence, we fixed the C-concentration to $3.8 \cdot 10^{19}$ cm$^{-3}$ while the thickness of the Si:C layers was set to 0.54 nm, 0.9 nm, 2.3 nm, 4.5 nm, 9 nm, 23 nm, 45 nm, and 82 nm. The buffer and cap thicknesses were adapted accordingly to maintain the same overall epilayer thickness.

**Photoluminescence characterization**

To perform microphotoluminescence (μ-PL) measurements at low temperatures (5 K), the samples were glued to the coldfinger of a liquid-helium (LHe) flow cryostat. For excitation, we used a continuous-wave (cw) diode-pumped solid-state (DPSS) laser emitting at 473 nm and a laser power of max. 6 mW (measured below the cryostat window). The laser was focused, and the luminescence signal was collected via an infinity-corrected microscope objective with 0.26 numerical aperture (NA) for the ensemble measurements. The μ-PL spectra were recorded via a 500 mm focal-length Czerny-Turner spectrometer with three interchangeable ruled gratings (100, 300, and 900 l/mm) connected to a liquid-nitrogen (LN2) cooled 1024 pixel InGaAs line detector.

*Ab initio* **calculation methods**

The considered defects were modeled by supercell plane wave density functional theory (DFT) as implemented in VASP.[54-56] We used 512-atom (4×4×4 multiple of the conventional Bravais-cell), and all atoms were allowed to relax in a constant volume till the forces were below 0.01 eV/Å. The $\Gamma$-point approximation was used for Brillouin-zone sampling. The lattice constant was taken from our earlier Heyd-Scuzeria-Ernzerhof HSE06 work to be 5.4307 Å (in good agreement with experiment.[57] The zero phonon line (ZPL) was obtained as the energy difference of the relaxed ground and excited states, the latter calculated with constrained occupation or $\Delta$SCF method. To obtain the correct ZPL for the singlet-to-singlet transition, an exchange correction was applied (e.g., Ref. [11]). The spectrum of the phonon replicas were computed by the generating function method,[58] based on vibration calculations using the Perdew-Burke-Ernzerhof (PBE) functional.[59]

**Acknowledgments**

We gratefully acknowledge funding support from the Austrian Science Funds (FWF) through projects Y1238-N36 and P36608. Support of accelerator operation at Uppsala University by the Swedish Research Council VR-RFI (Contracts No. 2017-00646_9 and No. 2019_00191) and the Swedish Foundation for Strategic Research (Contract No. RIF14-0053) is gratefully acknowledged. The support from European Commission for the EIC Pathfinder project QuMicro (Grant No. 101046911) is acknowledged.

**Correspondence and requests for materials** should be addressed to Johannes Aberl and Moritz Brehm.

# Supplementary material:

# All-epitaxial self-assembly of silicon color centers confined within atomically thin two-dimensional layers using ultra-low temperature epitaxy

Johannes Aberl[1,*], Enrique Prado Navarrete[1], Merve Karaman[1], Diego Haya Enriquez[1], Christoph Wilflingseder[1], Andreas Salomon[1], Daniel Primetzhofer[2], Markus Andreas Schubert[3], Giovanni Capellini[3,4], Thomas Fromherz[1], Peter Deák[5], Péter Udvarhelyi[5,6], Li Song[5], Ádám Gali[5,6,7], Moritz Brehm[1,*]

[1] *Institute of Semiconductor and Solid State Physics, Johannes Kepler University, Altenberger Straße 69, Linz 4040, Austria*

[2] *Department of Physics and Astronomy, Uppsala University, Box 516, 75120 Uppsala, Sweden*

[3] *IHP—Leibniz-Institut für innovative Mikroelektronik, Frankfurt (Oder), Germany*

[4] *Dipartimento di Scienze, Universita degli Studi Roma Tre, Rome, Italy*

[5] *HUN-REN Wigner Research Centre for Physics, P.O. Box 49, H-1525, Budapest, Hungary*

[6] *Department of Atomic Physics, Institute of Physics, Budapest University of Technology and Economics, Műegyetem rakpart 3., Budapest H-1111, Hungary*

[7] *MTA-WFK Lendület "Momentum" Semiconductor Nanostructures Research Group, P.O. Box 49, H-1525, Budapest, Hungary*

\* Johannes.aberl@jku.at
\* moritz.brehm@jku.at

**Ultra-low temperature epitaxy**

Conventional wisdom suggests that epitaxy growth temperatures ($T_G$) above ~500°C are needed to keep point defect densities low and the epitaxial quality high. However, at typical Si $T_G$s of 500°C-700°C, Si-based quantum emitters, such as G-centers or W-centers, disintegrate within common growth times by recrystallization.[1,2] Maintaining the SiCC emitter properties requires reducing the $T_G$s far below conventional values, i.e., ULT growth as Si quantum emitters like G-centers could sustain a thermal budget of $T_G$s ~300°C for several tens of minutes.[1,2] Therefore, playing with the thermal budgets for SiCC formation and annihilation, ULT growth can, in principle, allow for the formation of SiCCs due to kinetic limitations while allowing for overgrowth with high-quality Si (see Fig. 2).

**Importance of the growth pressure**

Typically, for MBE growth, a growth pressure ($P_G$) of $10^{-8}$-$10^{-9}$ mbar (at base pressures around $1 \cdot 10^{-10}$ mbar) has been considered sufficient but leads to a rate of gas impingement on the substrate of $10^{-2}$ - $10^{-3}$ monolayers (ML) per second, see inset of Fig. S1a.[3] But at high $T_G$, the sticking coefficients are small, and the gas can desorb from the Si surface.[3] However, ULT growth is less forgiving, as impurities *will* stick to the layer.[4] Thus, at a growth rate of 0.1 Å/s, one impurity per 10-100 Si or Ge atoms could be incorporated. Increasing the growth rate can be beneficial, but clearly, the main knob to turn is reducing $P_G$ to a minimum. To keep these impurities at "intrinsic semiconductor" levels, i.e., below $10^{16}$ cm$^{-3}$, less than one foreign atom per $10^6$ crystal atoms must be incorporated. To achieve these conditions, we have applied extensive and prolonged chamber conditioning, degassing, pumping, and gettering strategies to reduce $P_G$ during MBE growth deep into the ultra-high-vacuum range (see Fig. S1). For our growth rates, a ratio of growth rate vs rate of gas impingement reaches $10^5$ to $10^6$ or higher for $P_G \leq 5 \cdot 10^{-11}$ mbar, see Fig. S1. Even for highly C-doped Si layers for which the C source is heated up to >1500°C, $P_G$ remains at ~$10^{-10}$ mbar, see Fig. S1b.

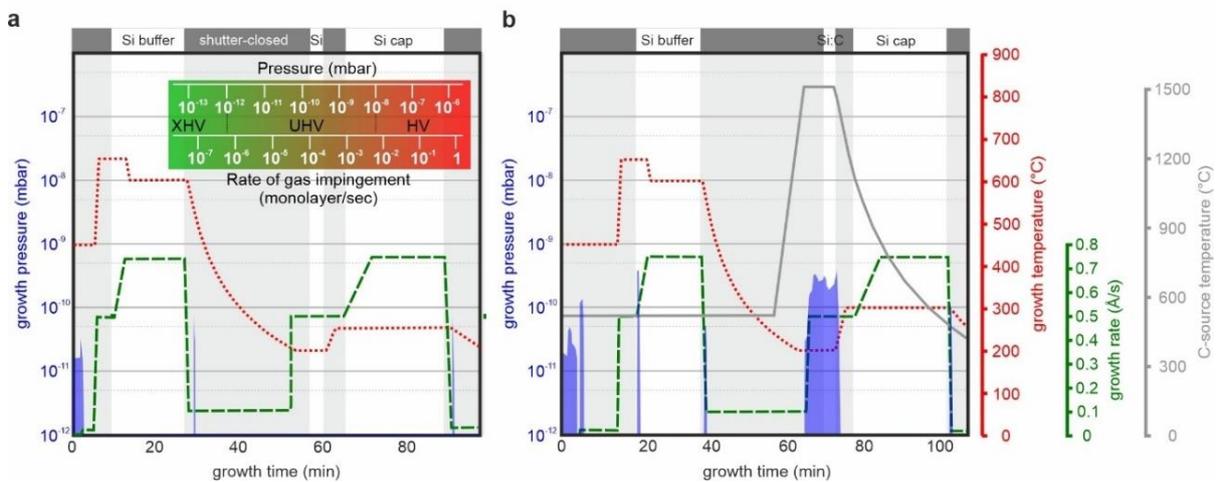

***Fig. S1:*** *Untypically-low growth pressures are needed for ULT growth of deterministic layer quality. Growth parameters versus growth time: Growth pressure (blue areas), substrate temperature (red dotted lines), Si growth rate (green dashed lines), and carbon source temperature (grey curve).* ***a*** *Growth parameters for W-center formation within a 9 nm thin layer, grown at 200°C and overgrown at 250°C.* ***b*** *Growth parameters for G'-center formation within a 9 nm thin Si layer doped with C ($3.8 \cdot 10^{19}$ cm$^{-3}$). Inset in **a**: Relationship between growth pressure and rage of residual gas impingement, essential for epitaxy growth at temperatures <350°C. Within the gray shaded areas, the growth temperature was ramped to its setpoint with all effusion cells and evaporators closed.*

**MBE versus CVD**

For forming self-assembled color centers, epitaxy methods like MBE are necessary. The standard in Si-related industries is chemical vapor deposition (CVD), also because many wafers can be overgrown in parallel by CVD. MBE, however, is versatile and exceptionally flexible, and many advances in semiconductor technology have thus been based on MBE. The particular key feature of MBE is the complete decoupling of $T_G$ and the growth rates of atoms impinging on the substrate. This is not the case for CVD, which requires the chemical decomposition of precursor gases that only occurs at elevated $T_G$s. For the creation of Si-based color centers, the game changer in favor of MBE is the accessibility of the ULT regime ($T_G$=100°C-300°C). For CVD, decreasing $T_G$ for epitaxy results in a decrease in the growth rate until the processes become ineffective. Regular precursors only provide reasonable growth rates of a few nm/min down to $T_G$~500°C for $SiH_4$ and ~350°C for $GeH_4$.[5] For highly epitaxial Si layers, necessary growth temperatures of ≤300°C cannot be reached by epitaxy methods that rely on the decomposition of precursor gases.

We note that MBE is used in industry, e.g., for the epitaxial growth of vertical cavity surface emitting lasers (VCSELs), and coupled CVD - MBE systems are employed to grow III-V quantum dot lasers on Si for Si photonics purposes. Thereby, the relaxed buffers are grown by CVD and the active emitters by MBE. Following the needs of the industry, a high-capacity MBE production tool (MBE 8000), able to handle four 200 mm wafers, has become recently available through a joint development program of Riber SA and IntelliEPI.[6]

**TEM analysis of the Si capping layer:**

From the PL spectra presented in Fig. 2 of the main text, we found that a $T_{cap}$ of 300°C minimizes the signal from here unwanted point defect emission centers like W or T-centers while maintaining the emission intensity of the underlying G'-center layer. Here, we additionally investigated the structural quality of the Si capping layer grown at $T_{cap}$ = 300°C by high-resolution cross-sectional transmission electron microscopy. Preparation for TEM was done conventionally by grinding, polishing, and Ar ion thinning. We have used a FEI Tecnai Osiris operated at an acceleration voltage of 200 kV for TEM investigation.

Figure S2 presents the overview image and a high-resolution section. The contrast between the high-temperature Si buffer, the low-temperature Si:C (9 nm thickness and C-concentration of $3.8 \cdot 10^{19}$ cm$^{-3}$), and the Si capping layer is too weak to see clear interfaces. The high-resolution image on the right side of Fig. S2, highlights the excellent crystalline quality of the Si capping layer, despite the low growth temperature of 300°C.

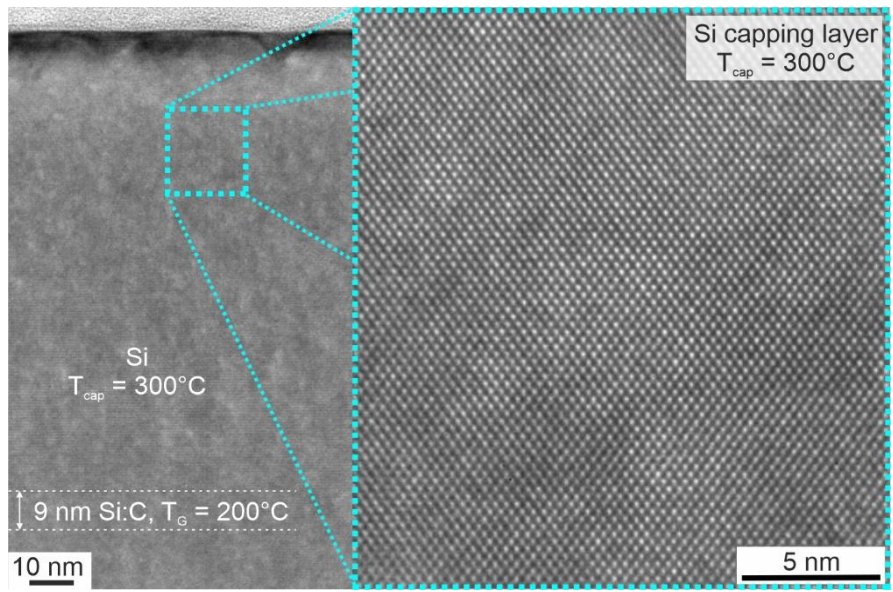

***Fig. S2:*** *Cross-sectional transmission electron microscopy images of a 9 nm thick Si:C layer (C-concentration of $3.8 \cdot 10^{19}$ cm$^{-3}$) capped with crystalline Si grown at $T_{cap}$ = 300°C.*

**Comparison of the observed SiCC emitter type with traditional G-centers:**

In Figure S3a, we compare the PL emission spectrum of the investigated G'-centers to traditional G-centers fabricated via C-ion implantation into epitaxial Si layers. For this purpose, a 300 nm thick Si epilayer was grown on a full 4 inch FZ Si(001) wafer using a Si growth rate of 0.9 Å/s. During growth of the first 40 nm, $T_G$ was linearly ramped from 475°C to 700°C while the remaining 260 nm were grown at a constant $T_G$ = 700°C. After growth, the wafer has been annealed *in-situ* at 750°C for 1 hour. For the creation of high-density common G-centers, the wafer has been irradiated with $5·10^{14}$ cm$^{-2}$ $^{12}$C$^+$ ions under a tilt angle of 7° using an ion energy of 34.5 keV. According to SRIM/TRIM simulations, the latter corresponds to a mean projected ion range of ~110 nm. The ion implantation was performed at room temperature w/o any post-implantation annealing. For the comparative PL measurements, 4 x 4 mm pieces have been cut from the wafer center followed by a (pre)cleaning procedure similar to that described in the methods section of the main article.

We observe the typical spectral fingerprint of the "traditional" G-center ensemble[7], i.e. a bright zero-phonon line (ZPL) at ~1278.5 nm with an ensemble linewidth of 0.57 nm at FWHM, an extended phonon side band with a well-known local phonon mode(s) (LPM, also called E-line) at ~1381 nm (and ~1500.8 nm) and a Debye-Waller factor of up to 17.5% at 5K. From the comparison (c.f. Fig. S3a) with the G' centers obtained/described/created in this work, the qualitative similarities between the PL spectra of the two emitter types become apparent, while all spectral features of the G' emitters are shifted to longer wavelengths as compared to traditional G-centers.

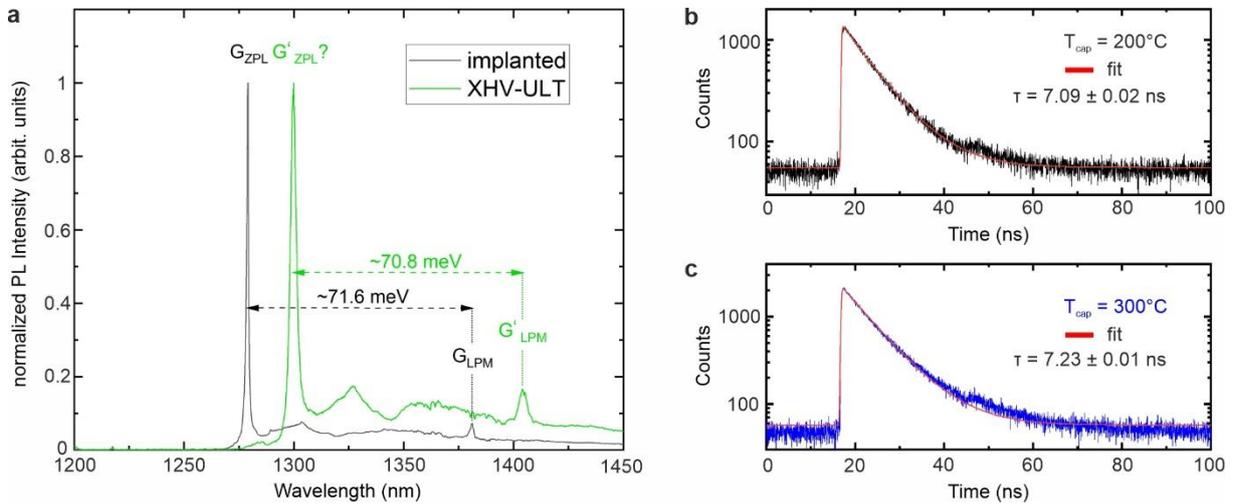

*Fig. S3: a* Normalized PL spectra of the G' (green), observed after Si color center self-assembly with a high C-concentration of $3.8·10^{19}$ cm$^{-3}$, and the traditional G-center (gray), observed after $^{12}$C$^+$ ion implantation. The respective zero-phonon lines ($G_{ZPL}$ and $G'_{ZPL}$) as well as the local phonon modes ($G_{LPM}$ and $G'_{LPM}$) are indicated. *b, c* Distribution of the PL photon emission times of a G'-center into its ZPL after pulsed excitation by a diode laser emitting at 442 nm. From fitting the observed time distribution to a single exponential decay superimposed on a noise background, excited state lifetimes of about 7 ns are obtained for both the sample with a Si cap grown at *b* $T_{cap}$ = 200°C and *c* 300°C.

**Lifetime measurements of G'-centers**

Figures S3b and S3c depict the results of excited state lifetime measurements of G' emitter ensembles by time-correlated single photon counting (TCSPC) experiments. The respective samples consist of a 9 nm thick Si:C layer, doped to a C-concentration of $3.8·10^{19}$ cm$^{-3}$, and capped at two different capping layer growth temperatures, $T_{cap}$ = 200°C (Fig. S3b) and $T_{cap}$ = 300°C (Fig. S3c). In both cases, the statistics of the time lag of ZPL photon emission after the excitation pulse were fitted with a single exponential function superimposed on a noise background, and the obtained excited state lifetimes amounted to $\tau \approx 7$ ns. For these measurements, the samples were excited by a pulsed single-mode diode laser (wavelength of 442 nm), with a pulse width of less than 200 ps at a repetition rate of 10 MHz and an average optical power ranging of 440 µW. To select the ZPL part of the resulting PL, the emission was spectrally filtered using a band-pass filter with a central wavelength of 1300 nm and a full-width-at-half-maximum of ~12 nm. The residual signal was further coupled to a single-mode fiber connected to a Single Quantum superconducting nanowire single photon detector (SNSPD) operated at 1.8 K. A PicoHarp 300 time tagging electronics operated in time-correlated photon counting (TCSPC) mode was then used to obtain the time-resolved PL decay.

*Ab initio* **results**

We employed *ab initio* calculations to identify the G'-center and reveal its potential for quantum technologies. We assumed, based on the similarities between the features of the PL spectra of the G-center and G'-center, that the core of the defect is the atomic structure of the G-center but a nearby carbon impurity or Si self-interstitial may perturb it. The possible models are carbon substitutional ($C_{Si}$), carbon interstitial ($C_i$) or silicon interstitial ($Si_i$) near $C_{Si}$-$Si_i$-$C_{Si}$ defect (so-called B-form of G-center that we simply label by 'G' for the sake of simplicity) that are depicted in Fig. S4.

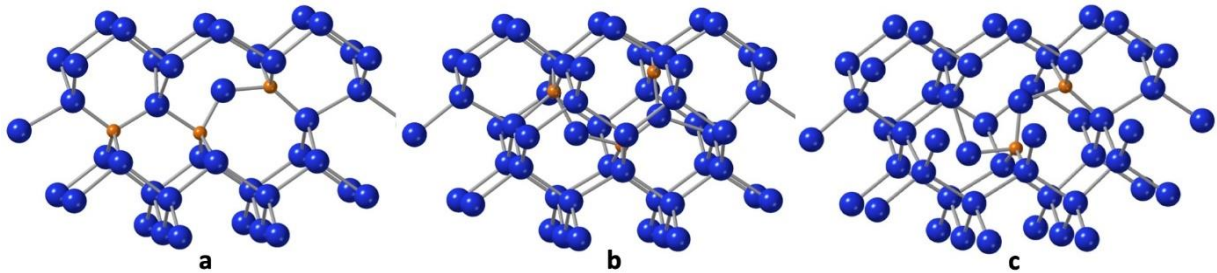

***Fig. S4:*** *Considered models of the G'-center.* ***a*** *$C_{Si}$+G,* ***b*** *$C_i$+G and* ***c*** *$Si_i$+G defects.*

The electronic structure of $C_i$+G and $Si_i$+G defects exhibits additional in-gap defect levels when compared to that of G-center (c.f., Fig. S5). On the other hand, the electronic structure of $C_{Si}$+G defect is only a small perturbation to the G-center's.

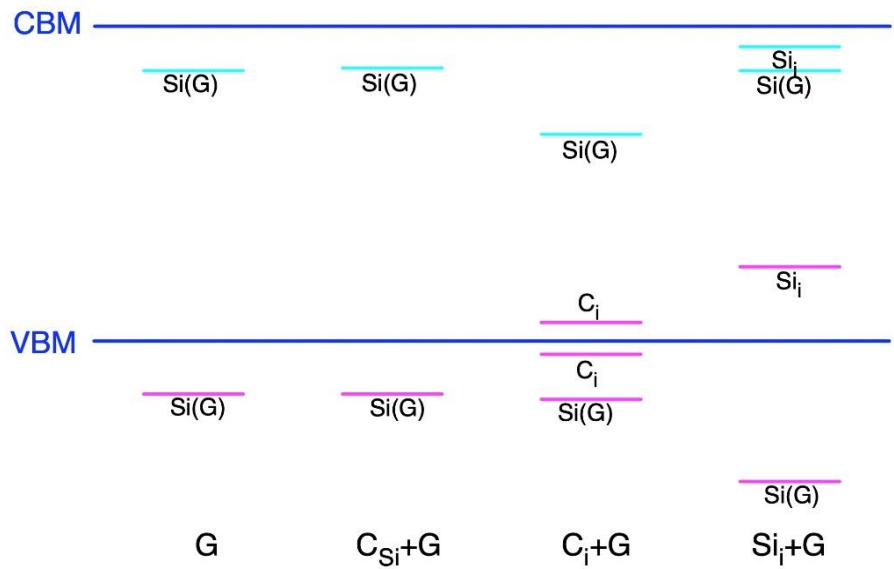

***Fig. S5:*** *HSE06 Kohn-Sham levels of the states localized on defect atoms in the G-center and in the three models of the G'-center. The cyan and magenta-colored defect levels are empty and fully occupied, respectively. CBM: conduction band minimum. VBM: valence band maximum. The calculated band gap is 1.16 eV. Localization of these states are labeled as follows: Si(G): dangling bond of silicon interstitial*

*in the $C_{Si}$-$Si_i$-$C_{Si}$ unit; $C_i$: dangling bond of carbon interstitial; $Si_i$: dangling bond of silicon self-interstitial near the $C_{Si}$-$Si_i$-$C_{Si}$ unit.*

The calculated HSE06 ZPL energies are at 920 and 934 meV for $C_{Si}$+G defect and G-center, respectively, so 14 meV red-shift appears when the ZPL of G-center is used as a reference. We note that we do not apply here local Hubbard-type correction to the p-orbital of $Si_i$ in the $C_{Si}$-$Si_i$-$C_{Si}$ unit in contrast to our previous work (Ref. [11] in the main text) because that correction might be defect specific that would prohibit the direct comparison of defects' properties. In our experiment, the G'-center has 17 meV red-shift in the ZPL energy. The calculated local vibration mode associated with the $C_{Si}$ motions in the $C_{Si}$-$Si_i$-$C_{Si}$ chain [Fig. S6(a)] that is a dominant peak in the phonon sideband of the PL spectra at 70.8 meV and 71.6 meV for the G'-center and G-center, respectively [Fig. S3(a)] appears at 68.7 meV and 69.9 meV [Fig. S6(b)] where the difference between these peaks in the experimental and simulated spectra is at 0.8 meV and 1.2 meV, respectively. The simulated PL phonon sideband agrees well with those in experiments where the prominent peak appears. Based on these results we assign G'-center to the $C_{Si}$+G defect.

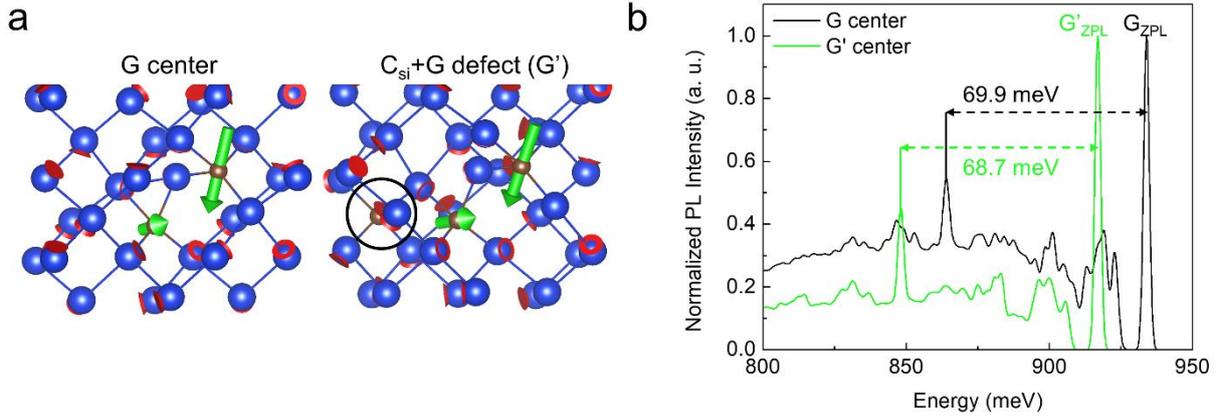

*Fig. S6: Prominent local vibration mode in the G-center and $C_{Si}$+G defect (G'-center). **a** The geometry of the defects is depcted where the black circle indicates the location of $C_{Si}$. The arrows show the direction and the amplitude of vibrating ions that are mainly localized on the two carbon atoms near the Si interstitial. **b** The simulated PL spectrum of the G-center and $C_{Si}$+G defect (G'-center) with the prominent local vibration modes as depicted in **a**.*

We also find that the G'-center has a triplet level at 0.63 eV above the ground state. The position of this metastable triplet level is reminiscent to the position of the metastable triplet level in the G-center (0.67 eV). In our previous work (Ref. [11] in the main text) we showed that this metastable triplet state was observed in the optically detected magnetic resonance (ODMR) of the G-center. Since the level structures of G-center (Ref. 11) and G'-center are very similar it is highly plausible that G'-center experiences the same ODMR effect. Although, the exact microscopic mechanism behind the ODMR effect for the G-center has not yet been revealed but the occurrence of the ODMR signal for the G-center clearly indicates that optical spin-polarization and optical readout of the electron spin in the metastable triplet state are doable. We propose that the same mechanism should occur in the G'-center too, so the same optical spin-polarization and readout techniques can be applied for the G'-center. We argue that our experiments realize a precise control of the location and density of the G'- center, which has become a very promising candidate for realizing spin-to-photon interface with emission in the telecom O-band.